\documentclass[aps,prl,twocolumn,superscriptaddress]{revtex4}

\usepackage{amssymb}		
\usepackage{amsmath}    
\usepackage{graphicx}   
\usepackage{amsfonts}
\usepackage{subfigure}  
\usepackage{hyperref}   
\usepackage{epstopdf}
\usepackage[version=3]{mhchem}		

\newcommand{\abs}[1]{\left\vert#1\right\vert}

\begin{document}

\title{Terahertz electron-hole recollisions in GaAs/AlGaAs quantum wells: robustness to scattering by optical phonons and thermal fluctuations}

\author{Hunter Banks}
\author{Ben Zaks}
\affiliation{Physics Department, University of California, Santa Barbara, USA}
\affiliation{Institute for Terahertz Science and Technology, University of California, Santa Barbara, USA}
\author{Fan Yang}
\affiliation{Department of Physics, The Chinese University of Hong Kong, Hong Kong, China}
\author{Shawn Mack}
\author{Arthur C. Gossard}
\affiliation{Materials Department, University of California, Santa Barbara, USA}
\author{Renbao Liu}
\affiliation{Department of Physics, The Chinese University of Hong Kong, Hong Kong, China}
\author{Mark S. Sherwin}
\affiliation{Physics Department, University of California, Santa Barbara, USA}
\affiliation{Institute for Terahertz Science and Technology, University of California, Santa Barbara, USA}

\date{\today}

\begin{abstract}
Electron-hole recollisions are induced by resonantly injecting excitons with a near-IR laser at frequency $f_{\text{NIR}}$ into quantum wells driven by a ~10 kV/cm field oscillating at $f_{\text{THz}} = 0.57$ THz.   At $T=12$ K, up to 18 sidebands are observed at frequencies $f_{\text{sideband}}=f_{\text{NIR}}+2n f_{\text{THz}}$, with $-8 \le 2n \le 28$.  Electrons and holes recollide with total kinetic energies up to 57 meV, well above the $E_{\text{LO}} = 36$ meV threshold for longitudinal optical (LO) phonon emission.  Sidebands with order up to $2n=22$ persist up to room temperature.  A simple model shows that LO phonon scattering suppresses but does not eliminate sidebands associated with kinetic energies above $E_{\text{LO}}$.
\end{abstract}


\maketitle

	\indent The interaction of electrons and holes in semiconductors has long been a rich area of study in physics.  In most cases, the electron and hole are treated as point particles within the effective mass approximation \cite{Luttinger:1955vx}.  In this approximation, the effects of the lattice are parametrized by the dielectric constant of the semiconductor and the effective masses of the electrons and holes.  Calculations based on the effective mass approximation successfully predict, for example, the binding energies and the frequencies of internal transitions of impurity-bound electrons and excitons in GaAs \cite{Yu:2010yu} and in GaAs/AlGaAs quantum wells \cite{Greene:1984tv,Cerne:1996tm,Salib:1996tc}.  The effective mass approximation, however, belies the rich complexity of the microscopic physics.  In GaAs, for example, ground-state excitons are collective excitations of more than $10^5$ atoms in the crystal.\\
	\indent The advent of sources of intense terahertz electromagnetic radiation enables the study of semiconductors in a regime where  time-dependent perturbation theory fails completely. Exciting new quantum coherent phenomena emerge, including the dynamical Franz-Keldysh effect \cite{Nordstrom:1998ab,Shinokita:2010dj}, non-linear excitonic effects \cite{Carter:2005vy,Danielson:2007dn,Wagner:2010eq,Ewers:2012kz}, and high-order sideband generation \cite{Yan:2008js,Zaks:2012fp,Zaks:2013jg}.  High-order sideband generation is a cousin of high-order harmonic generation from atoms in intense laser fields \cite{Krause:1992ug,Lewenstein:1994wf,Corkum:2007hb}.  Excitons are created by a near-infrared laser in a semiconductor driven by an intense terahertz field.  The terahertz field ionizes the exciton into an electron and a hole that it then pulls apart and smashes back together.  Upon recollision, the electron and hole can recombine across the band gap with the acquired kinetic energy carried off by light with frequency above that of the near-IR laser. \\
	\indent In this Letter, we report that the electron-hole recollision process---which, in its theoretical descriptions to date has been treated as a purely ballistic process---is surprisingly robust against scattering.  We report electron-hole recollisions both with kinetic energy well above the threshold for emission of a longitudinal optical (LO) phonon and at room temperature.  We model our results by treating the electron-hole pair as a single particle with the appropriate reduced mass subject to dephasing and LO phonon scattering.\\
	\begin{figure}[b!]
		\includegraphics{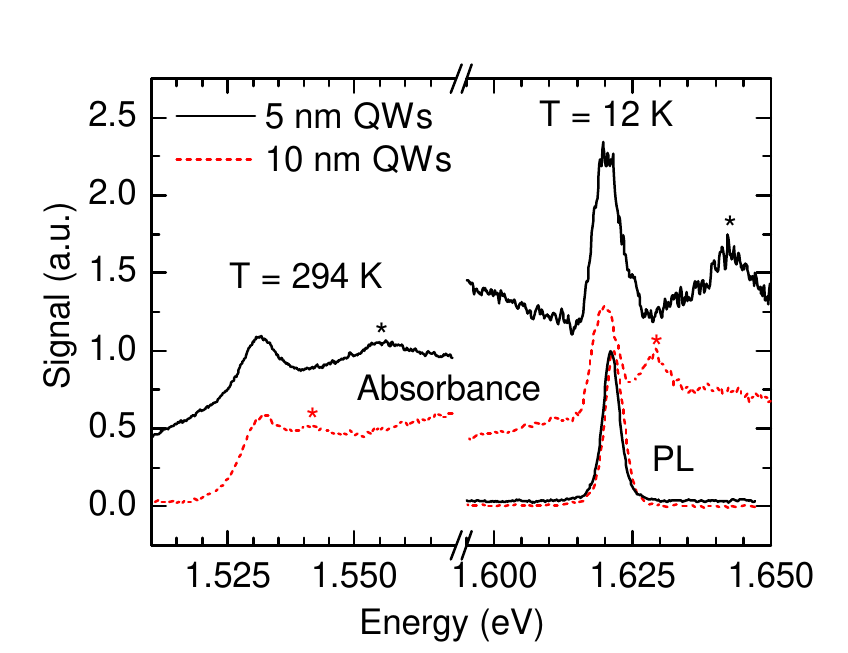}
		\caption{PL and absorption measurements for $T = 12 \text{ K}$ and $T = 294 \text{ K}$.  The absorption spectra are offset for clarity. The heavy hole peaks in both measurements at low temperature agree very well.  The second, minor peaks denoted with asterisks in the absorption spectra are assigned to the light hole exciton.}
		\label{fig1}
	\end{figure}
	\begin{figure*}
		\includegraphics[scale=1]{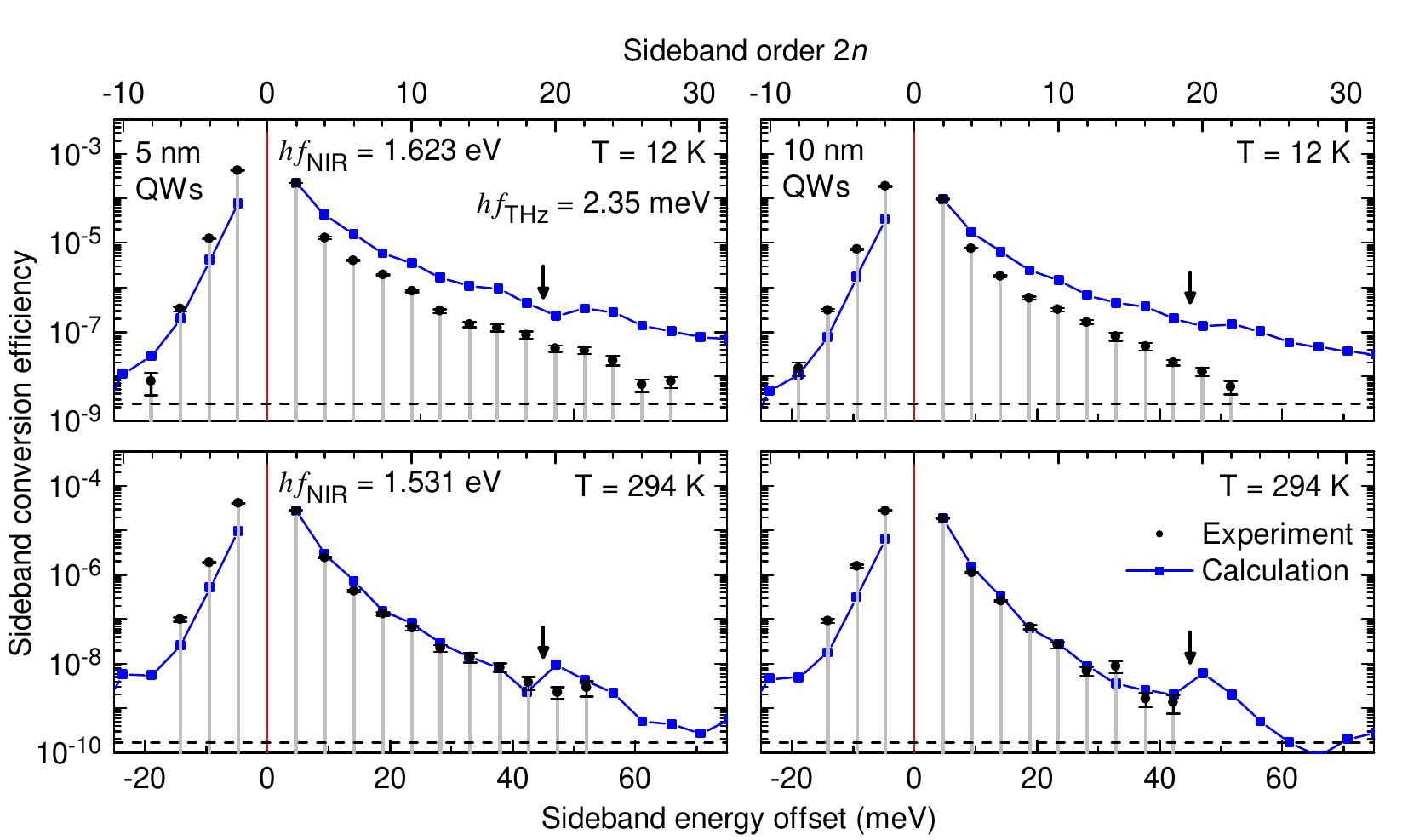}
		\caption{Measured high-order sideband strengths compared with theoretical calculations.  The observed sideband conversion efficiencies versus sideband energy offset $(= hf_{\text{sideband}} - hf_{\text{NIR}})$ and sideband order for the 5 nm QW sample at $T = 12$ K (top left) and $T = 294$ K (bottom left) and for the 10 nm QW sample at $T = 12$ K (top right) and $T = 294$ K (bottom right) are plotted as black circles. For each sideband, the conversion efficiency is calculated by integrating the individual sideband peak and then dividing by the photomultiplier tube's response to the NIR laser with no sample. Calculated sideband strengths, normalized to the conversion efficiency for the $n=1$ sideband, are plotted as blue squares connected by a line to guide the eye.  The vertical red lines at 0 meV denote $hf_{\text{NIR}}$.  The dashed line represents the noise floor of detection.  The arrows indicate the threshold at which, within the three step model \cite{Corkum:1993vt}, the total kinetic energy of the colliding electron and hole is larger than the longitudinal optical phonon energy.  }
		\label{fig2}
	\end{figure*}
	\indent Two quantum well (QW) samples with 20 QW repetitions were studied.  Both samples were grown on semi-insulating GaAs substrates by molecular beam epitaxy with no intentional doping.  The 5 nm QW sample consists of 5 nm GaAs QWs with 20 nm \ce{Al_{0.3}Ga_{0.7}As} barriers and the 10 nm QW sample consists of 10 nm \ce{Al_{0.05}Ga_{0.95}As} QWs with 20 nm \ce{Al_{0.3}Ga_{0.7}As} barriers.  For the experimental convenience of matching excitonic absorption peaks in the different samples, the band gap of the wider well was raised by adding 5\% aluminum to match the 2D band gap of the narrower well.  Because HSG is measured in transmission and the exciton absorption lines of the QWs are above the band gap of bulk GaAs, the substrate was removed by a series of etches \cite{Moon:1998vx, Chang:2001tf}.  The final result of the processing was a 1 $\times$ 2 mm membrane less than 1 $\mu$m thick.  Details of the growth, processing and phase matching considerations can be found in the Supplemental Material.\\
	\indent The samples have nearly identical heavy hole exciton energies, as designed.  At $T=12$ K, spectra of the absorbance (proportional to the density of states) and photoluminescence (which reveals the lowest-energy radiative state) show a strong peak at 1.62 eV associated with the heavy-hole exciton in both samples (see Fig. \ref{fig1}).  An excitonic peak persists in the absorbance spectrum up to room temperature but is red-shifted by 100 meV because of the temperature dependence of the band gap.  Absorption features associated with light-hole excitons are present above the heavy-hole exciton.  Because of greater confinement the light-hole exciton is at higher energy in the 5 nm QW sample.  \\
	\indent In order to generate sidebands, we simultaneously illuminated the sample with the NIR laser tuned to the heavy hole exciton and intense 0.57 THz radiation from the UCSB FEL.  The radiation from both sources was linearly copolarized and focused onto the same spot.  Radii were approximately 100 $\mu$m and 1 mm for the NIR and THz spots, respectively.  The strength of the terahertz electric field was 11.5 kV/cm within the quantum wells.  The power of the NIR laser was 39 mW (60 mW) at $T = 12$ K (294 K). The transmitted NIR radiation was dispersed in a double monochromator and detected by a photomultiplier tube. Further details can be found in Ref. \cite{Zaks:2012fp}.\\
		\indent In the 5 nm QW sample at 12 K, we observed up to 4 sidebands below ($-8$th order) and 14 sidebands above (28th order) the near-IR laser (See Fig. \ref{fig2}, top left).  The intensity of the sidebands with negative order decreases exponentially with $n$.  The intensity of the sidebands with positive order decreases much more gently with no clear systematic dependence on $n$.  We observed a similar spectrum of sidebands in the 10 nm QW sample.  However, the intensity of each sideband is approximately three times smaller than the corresponding sideband in the 5 nm QW sample, and the highest sideband is 22nd order rather than 28th.  \\
	\indent Remarkably, HSG persists to room temperature in both samples, with sideband orders as high as 22nd and 18th for the 5 nm and 10 nm QW samples, respectively.  The sideband conversion efficiencies are approximately one order of magnitude smaller at 294 K compared to 12 K, but otherwise the HSG spectra have similar shapes.  \\
	\indent Many features of the data can be explained qualitatively by adapting the three-step model for high-order harmonic generation (HHG) from atoms in intense laser fields to the phenomenon of HSG.  Step zero, which is not present in the case of HHG for atoms, is the creation of the electron-hole pairs by the NIR laser at a particular phase of the THz field.  For these experiments, the NIR laser is tuned to the exciton resonance stemming from the highest heavy hole subband and the lowest electron subband, thus creating carriers with in-plane masses of $m_e^* = 0.067 m_e$ and $m_h^* = 0.103 m_e$ for the electron and hole, respectively (The hole mass is appropriate for a 5 nm QW within the Kohn-Luttinger model \cite{Luttinger:1955vx}).  Figure \ref{fig3} illustrates the next three steps that lead to the emission of the 28th order sideband.  In the first step, the exciton tunnel-ionizes in the strong THz field at a phase 54$^{\text{o}}$ after the maximum of the terahertz field (within the three-step model, the phase at ionization determines the kinetic energy at recollision, and hence the sideband energy offset---see the Supplemental Material and Ref. \cite{Corkum:1993vt}).  At this point, the carriers are assumed to be point particles with zero initial momentum.  The Keldysh parameter, $\gamma = \omega_{\text{THz}}\sqrt{2 \mu^* E_{\text{bind}}} / \left(e F_{\text{THz}} \right)$, the ratio of the average tunneling time to the driving field's period, is 0.2, well within the regime of tunnel ionization \cite{Keldysh:1965vh,Ilkov:1992wi}.  Here, $\mu^*$ ($= 0.041 m_e$) is the reduced mass of the electron-hole pair, $E_{\text{bind}}$ is the exciton binding energy, $e$ is the electric charge, and $\omega_{\text{THz}}$ and $F_{\text{THz}}$ are the angular frequency and electric field strength of the terahertz radiation, respectively.  In the second step, the electron and hole are assumed to be independent and to accelerate in the THz field classically.  In the third step, the electron and hole recollide to emit an NIR photon with larger energy than the photon that created them. \\
	\indent The high-order sidebands with positive $n$ can be analyzed to extract the kinetic energy of the electron-hole system at recollision.  The sideband energy offset is given by $E_{\text{SB}} = E_{K,\text{electron}} + E_{K,\text{hole}} + \Delta$, where $E_{K,i}$ is the kinetic energy for the $i$th particle and $\Delta$ is the detuning of the NIR laser from the 2D band gap.  Because this is an excitonic effect, $\Delta$ should be close to the binding energy of the exciton, approximately 9 meV for the 5 nm QW sample and 8 meV for the 10 nm QW \cite{Greene:1984tv}.  The 28th order sideband is 66 meV above the NIR laser energy.  Following this detuning assumption, the total kinetic energy of the electron and hole is $\sim 57$ meV.  This energy is significantly higher than the threshold for a charge carrier to emit a longitudinal optical (LO) phonon. Figure \ref{fig3} illustrates the classical trajectories (top) and kinetic energy (bottom) for the electron and hole associated with the 28th order sideband between ionization at $t=0$ and recollision.  The kinetic energy of the electron-hole system is above the LO phonon energy only for the final 50 fs before recollision.  Since this is below the typical electron-LO phonon scattering time of 100 fs, a large fraction of electron-hole pairs are likely to recollide before they emit a LO phonon---even though the electron-LO phonon scattering time is much shorter than the period of the terahertz radiation. This analysis suggests that the fundamentally ballistic process of HSG persists to room temperature because thermal fluctuations simply do not have enough time to disrupt electron-hole recollisions.  \\
	\indent Although there is no abrupt cutoff due to LO phonons, we observe far fewer sidebands than are predicted by the ballistic three-step model of HSG.   This model predicts a sharp cutoff for the sideband offset energy:  $\hbar \omega_{\text{cutoff}} = 3.17 U_{\text{P}} + E_{\text{bind}}$, where $U_{\text{P}}$ is the ponderomotive energy [$= e^2 F_{\text{THz}}^2 /(4 \mu^* \omega_{\text{THz}}^2)$].  Such a cutoff is observed for the harmonics in high-order harmonic generation experiments on atoms.  Inserting the parameters appropriate for our experiments, the ballistic three-step model predicts a cutoff near 65 sidebands (130th order), approximately 4 times higher than the highest-order sideband reported here.\\
	\begin{figure}
		\includegraphics{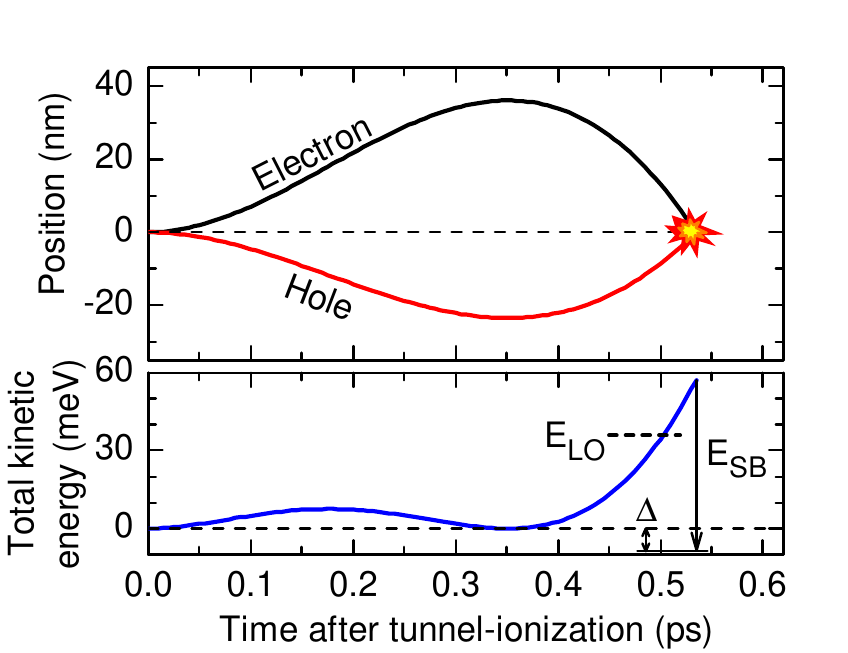}
		\caption{Classical calculation of electron-hole dynamics associated with 28th order sideband.  Top:  trajectories of the electron (black) and hole (red) show electron-hole separation reaches a maximum of 60 nm between tunnel ionization (0 ps) and recollision (flash).  Bottom:  total kinetic energy of the electron and hole. The total kinetic energy exceeds the threshold for LO phonon emission $E_{\text{LO}}$ (short dashed line) for only $\sim 50$ fs before emitting a sideband with sideband offset energy $E_{\text{SB}}$ (long vertical arrow) = Total kinetic energy + detuning $\Delta$ (double arrow).  The long dashed horizontal lines denote 0 nm (top) and 0 meV (bottom).  See text and supplementary materials for details.}
		\label{fig3}
	\end{figure}
	\indent To investigate this discrepancy, we have also performed the simplest calculation that captures the relevant quantum mechanical nature of the recollision process together with the most important scattering mechanisms, exciton dephasing and electron-LO phonon scattering. In this calculation, the intensity of the $2n$th order sideband is proportional to the integral in Eq. (1), which can be integrated numerically and is plotted in Figs. \ref{fig2} and \ref{fig4}.   
	\begin{align}
		P_{2n} &\propto \int e^{i S\left( \mathbf{k},t,\tau\right)}e^{-\gamma \tau}e^{-W \! \left(d,T,\mathbf{k}\right) \, \tau} \, \text{d} \mathbf{k} \, \text{d} t \, \text{d} \tau \\
		S \left(\mathbf{k}, t, \tau \right) &= - \int^t_{t - \tau} \frac{1}{2 \mu^*} [ \mathbf{k} + e \mathbf{A} \! \left( t'' \right) ]^2 \, \text{d} t'' - \Delta \tau + 2 n \omega_{\text{THz}} t.
	\end{align}
	Here, $S(\mathbf{k}, t, \tau)$  is the action, where $\mu^*$, $e$ and $\omega_{\text{THz}}$ were defined previously, $\mathbf{k}$ is the momentum (working in units where $\hbar=1$), and $\mathbf{A}$ is the vector potential of the oscillating terahertz field.   The exciton is created at $t-\tau$, the electron and hole evolve through $\tau$, and eventually annihilate at $t$, emitting sidebands with order $2n$.  The detuning $\Delta$ is assumed to be $E_{\text{bind}}$.  Eq. (1) represents an extension of the model used in Ref. \cite{Liu:2007ab} to include the exciton dephasing rate $\gamma$, which was estimated from the half-lidewidth of the heavy-hole absorption peak (2.8 meV for both samples at 12 K; 4 meV and 5 meV for the 5 nm and 10 nm QWs at 294 K, respectively), and the scattering rate due to emission or absorption of LO phonons $W(d, t, \mathbf{k})$.  $W$ is calculated from Ref. \cite{Constantinou:2001td} and depends on the QW width $d$, the temperature $T$, and the kinetic energy $\abs{\mathbf{k}}^2/(2\mu^*)$ of the carrier, which is assumed to be an electron with mass $\mu^*$.  In the calculation, $\omega_{\text{THz}} = 2 \pi \times 0.58$ THz (photon energy $E_{\text{THz}} = 2.4$ meV), $F_{\text{THz}} = 11$ kV/cm, and the photon energy of the NIR laser $E_{\text{NIR}} = 1.623$ eV.  The LO phonon energy is taken as 36.6 meV.  In order to isolate the effects of scattering on high-order sideband generation, many interactions, such as the Coulomb interaction and heavy hole-light hole mixing, have been neglected. \\
	\indent  These calculations refine the simple idea that HSG works ``too fast'' to be affected by LO phonons.  At low temperature, sideband intensities calculated with LO phonon scattering begin to fall below those calculated without such scattering very close to the threshold for LO phonon emission [see Fig. \ref{fig4}(a)].  Above this threshold, sideband intensities are suppressed but not eliminated.  At room temperature [Fig. \ref{fig4}(b)], all positive sidebands are suppressed because there is now a finite chance of absorbing an LO phonon from the thermal bath.  As at low temperatures, sidebands persist above the threshold for LO phonon emission.  
		\begin{figure}
			\includegraphics[scale=1]{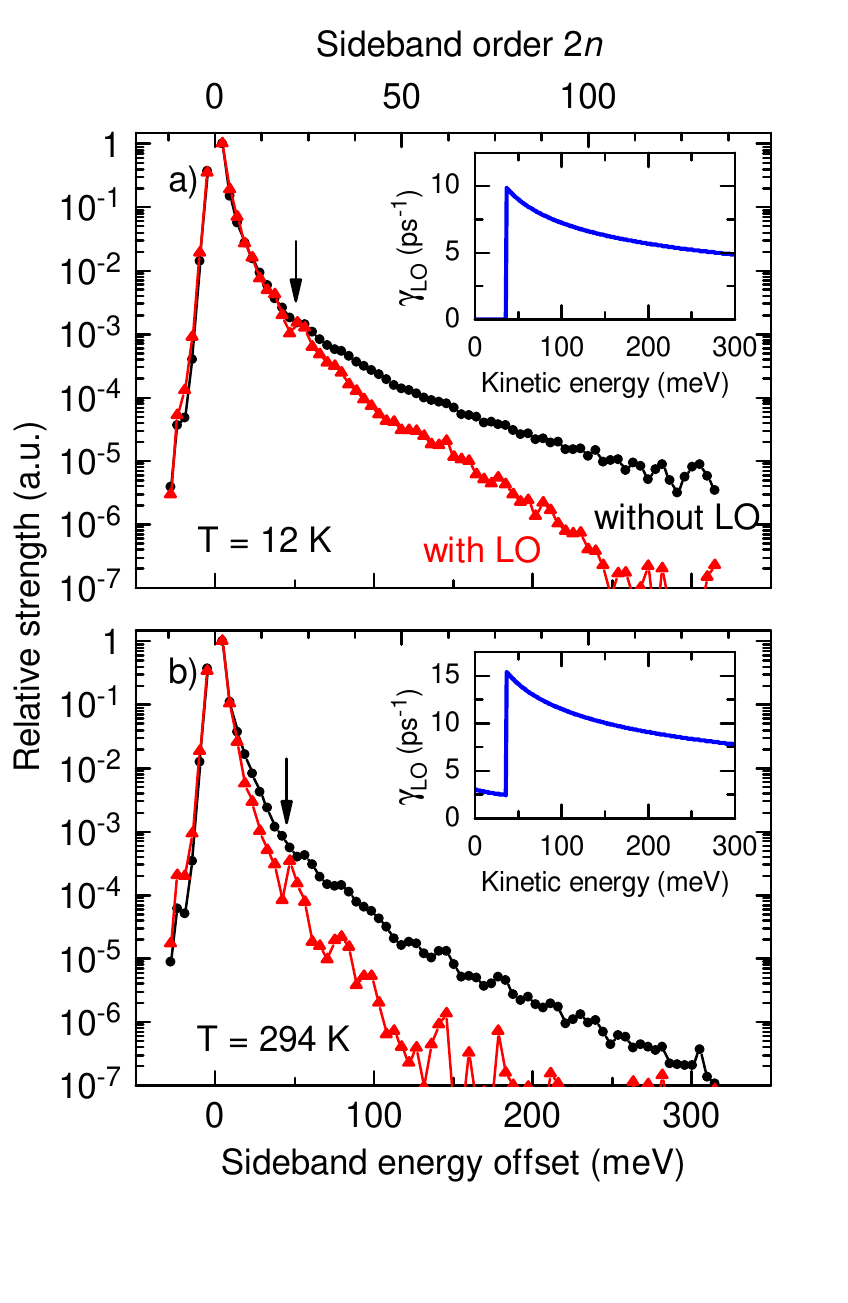}
			\caption{Calculations of the effects of longitudinal optical phonon scattering.  HSG calculations at (a) $T = 12$ K and (b) $T = 294$ K with (black circles) and without (red triangles) LO phonon scattering.  An arrow denotes the threshold for emitting a LO phonon in the 5 nm QW sample.  Excitonic dephasing is included in all calculations.  The inset is the calculated total energy-dependent scattering rate given by \cite{Constantinou:2001td}.}
			\label{fig4}
		\end{figure}\\
	\indent Calculations are compared with experimental data in Fig. \ref{fig2}.  At low temperature, the calculations overestimate the intensities of high-order sidebands for positive $n$ relative to the low-order positive sidebands.  At both temperatures, the calculations approximate the observed exponential decay with $n$ of the sidebands below the NIR laser line.  However, at room temperature, the calculated intensities of the positive sidebands agree remarkably well with experiment.  The agreement of experiment and theory at room temperature suggests that the decoherence at room temperature is so strong that it dominates the physics missing from the model.  \\
	\indent Our experimental and theoretical results suggest that suppressing scattering mechanisms may lead to the observation of a larger number of sidebands in future experiments.  One well-known method to decrease scattering by phonons is to reduce dimensionality \cite{Benisty:1991ub}.  Thus, experiments on HSG in quantum wires (1D) would be interesting to pursue, with the THz electric field polarized along the wire.  For this polarization, the electrons and holes can be driven arbitrarily far apart from one another, as in quantum wells and bulk semiconductors, the three-step model should still apply, and the ponderomotive energy will remain an important energy scale in the problem.  Reducing dimensionality further, to quantum dots (0D), results in completely different nonlinear physics, as the continuum is eliminated from the problem and the ponderomotive energy is no longer relevant.  However, even in the extreme case of a two-state atom, high-order harmonic generation has been predicted \cite{Sundaram:1990vm,FigueiradeMorissonFaria:2002cx} with a cutoff energy scale of $eF_{\text{THz}}a$, where $a$ is a length scale associated with the two-state atom.  Thus high-order sideband generation in quantum dots is an interesting possibility. \\
	\indent In conclusion, mixing strong $\sim 0.6$ THz terahertz radiation with a NIR laser beam has resulted in a comb of up to 19 frequencies spanning up to 21 THz (an approximately 30 nm range).  Observation of sidebands at room temperature relaxes a significant experimental restriction to HSG and opens up the scientific and the practical possibilities of this phenomenon. Both experiments and calculations show that electron-hole recollisions in semiconductors are surprisingly robust to, though still affected by, strong scattering processes that are not present in atoms but are unavoidable in the solid state. \\
	\indent Acknowledgments:  The authors thank David Enyeart for assistance operating the UCSB Free-Electron Laser.  This work was supported by NSF grant DMR-1006603 (H. B., B. Z. and M. S. S.), and Hong Kong RGC/GRF 401512 (F. Y. and R. B. L.).

\bibliography{HSG_bib}

\end{document}